\documentclass[preprint,aps,draft]{revtex4}
\usepackage[sumlimits]{amsmath}
\usepackage{latexsym}
\usepackage{amssymb}
\usepackage{euscript}
\usepackage{color}
\usepackage{graphicx}

\begin{document}
\def\be{\begin{equation}}
\def\ee{\end{equation}}
\def\bea{\begin{eqnarray}}
\def\eea{\end{eqnarray}}

\title{Geometrothermodynamics of higher dimensional black holes }

\author{Alessandro Bravetti$^{1,2}$, Davood Momeni$^3$, Ratbay Myrzakulov$^3$, and Hernando Quevedo$^{1,2,4}$}
\affiliation{$^1$Dipartimento di Fisica and ICRA, Universit\`a di Roma La Sapienza, Piazzale Aldo Moro 5, I-00185 Rome, Italy\\
$^2$Instituto de Ciencias Nucleares, Universidad Nacional Aut\'onoma de M\'exico, AP 70543, M\'exico DF 04510, Mexico \\
$^3$ Eurasian International Center for Theoretical Physics,
Eurasian National University, Astana 010008, Kazakhstan\\
$^4$Instituto de Cosmologia, Relatividade e Astrofisica ICRA - CBPF\\
Rua Dr. Xavier Sigaud, 150, CEP 22290-180, Rio de Janeiro, Brazil
}

\email{bravetti@icranet.org,d.momeni@yahoo.com,myrzakulov@gmail.ru,quevedo@nucleares.unam.mx}

\date{\today}

\begin{abstract}
We study the thermodynamics and geometrothermodynamics of different black hole configurations in 
more than four spacetime dimensions.  We use the response functions to 
find the conditions under which second order phase transitions occur  
in higher-dimensional static Reissner-Nordstr\"om and stationary Kerr black holes. 
Our results indicate that the equilibrium manifold of all these black hole configurations is in general curved and 
that curvature singularities appear exactly at those places where second order phase transitions occur.
\end{abstract}


\maketitle

\section{Introduction}

During the last century, differential geometry has become an essential element of theoretical physics. 
One of the most interesting examples of this fact is the application of Riemannian geometry in general relativity as a theory of the gravitational field.
Indeed, today we understand gravity as a manifestation of the Riemann curvature of spacetime so that measuring the curvature is equivalent to 
measuring the gravitational interaction. 
This is a consequence of the astonishing principle ``field strength = curvature", proposed originally by Einstein. 
Since the field strength can be considered as a measure of the gravitational interaction, it follows that 
that the entire idea of general relativity can be summarized in the principle ``interaction = curvature". 
The same principle is valid in the case of gauge theories. For instance, Maxwell's theory of electromagnetism 
can be described geometrically in terms of the elements of a principal fiber where 
the base manifold is the Minkowski spacetime, the standard fiber is the gauge group $U(1)$, which 
represents the internal symmetry of the electromagnetic interaction,
and the connection across the fibers is a local cross-section which takes values in the algebra of $U(1)$.   
If we replace the gauge group and the connection $U(1)$ by $SU(2)$ or $SU(3)$, we obtain the geometric description of 
the weak or the strong interaction, respectively. Although in the case of gauge theories the Riemannian curvature of
the base space vanishes, it is the gauge curvature of the principal fiber bundle which is equivalent to the interaction 
(see, for instance, \cite{frankel} for an introductory review).

Another important branch of theoretical physics is thermodynamics, and one may  wonder whether it is possible to 
represent it in the context of differential geometry. The first attempts in this direction were made in the 
pioneering works by Gibbs \cite{gibbs} and Caratheodory \cite{car} in which the language of differential forms were
introduced in thermodynamics. 
Riemannian geometry was first introduced in statistical physics and thermodynamics
by Rao \cite{rao45}, in 1945, by means of a metric whose components in local coordinates coincide with Fisher's
information matrix. Rao's original work has been followed up and extended by a number of
authors (see, e.g., \cite{amari85} for a review).  
Hessian metrics have been used intensively to study the geometry of the thermodynamics of
ordinary systems and black holes \cite{am,ama,aman06a,scws,caicho99,sst,med,mz,hernando2}.
An additional aspect of classical thermodynamics is that it is invariant with respect to 
Legendre transformations \cite{callen}, i.e., the properties of a given thermodynamic system
are independent of the choice of thermodynamic potential. In \cite{quev07}, 
this property was taken into account in the context of a  geometric description of thermodynamics.

It turns out that to correctly handle the Legendre transformations and the geometric version of the first 
law of thermodynamics, it is necessary to introduce a contact geometry  structure  in the thermodynamic phase space
\cite{her73}. The formalism of geometrothermodynamics (GTD) \cite{quev07} was recently proposed to unify in a consistent and Legendre invariant
manner the contact structure of the phase space with the Riemannian structure of the equilibrium space. As a result, 
Riemannian metrics are obtained for the equilibrium space that are no longer Hessian, and are 
invariant under Legendre transformations. One of the main goals of GTD is to interpret 
the curvature of the equilibrium space as a manifestation of the thermodynamic
interaction. This would imply that an ideal gas and its generalizations with no internal mechanic interaction
correspond to an equilibrium manifold with vanishing Riemannian curvature. In the case of interacting
systems with non-trivial structure of phase transitions, one would expect that the curvature is non-vanishing and
becomes singular near the points where phase transitions occur. This has been shown to be true in all the thermodynamic
systems investigated so far \cite{qstv10a}. In particular, all the black hole configurations of Einstein's theory 
in four dimensions were investigated in detail. In this work, we will analyze the geometric properties of the equilibrium manifold
of higher dimensional black holes. 

In recent years, black hole solutions in more than four spacetime dimensions have been the subject of intensive research. 
Extensions of general relativity to higher dimensional Riemannian spacetimes, provides more information about the fundamental properties of black holes. In dimensions higher than four, the uniqueness theorems do not hold due to the fact that there are  more possibilities to include more degrees of freedom. For example, in five dimensions, the additional rotational symmetry adds one more angular momentum to the rotating object. Different kinds of black objects have been found in higher dimensional spacetimes (see \cite{liv-rev} for a comprehensive review). Another interesting feature of higher dimensional black objects refers to the topology of the horizon. In four dimensional configurations, the topology of the Killing horizon is trivially fixed as $S^2$. But in five dimensions, we can have different topologies as $S^2\times S^1$ for black objects with ring singularities, and the string topology $S^2\times \mathcal{R}$ for black strings in supergravity extensions of  higher dimensional general relativity. Also, the phase transition structure of the black holes looks like completely different. 
The phase transition structure depends on the dimension of the spacetime. In the case of asymptotically anti-de Sitter black holes \cite{gpw11,roy11a,roy12,roy11b,qs08}, the black hole properties are strongly affected by the presence of the cosmological constant and, in some cases, the configuration becomes unstable \cite{konzhi09}.  
These different characteristic features of black objects in higher dimensional gravity theories (including vacuum general relativity, Einstein-Yang-Mills, etc.) encourage us to investigate the GTD formalism in this new area.

This work is organized as follows. In Sec. \ref{sec:gtd}, we review the main ingredients of the formalism of GTD, and present 
the thermodynamic metric that will be used in further sections to analyze the equilibrium manifold of black hole configurations.
In Secs. \ref{sec:rn} and \ref{sec:kerr}, we analyze the thermodynamic properties of the Reissner-Nordstr\"om and Kerr black holes
in dimensions higher than four, and present the explicit form of the metric for the corresponding equilibrium manifolds. 
 We show that the GTD analysis correctly reproduces the thermodynamic behavior of these configurations.
Finally, Sec. \ref{sec:con} contains discussions of our results and suggestions for further research.

\section{Geometrothermodynamics}
\label{sec:gtd}

To describe a  system with $n$ thermodynamic degrees of freedom it is convenient to introduce the equilibrium space ${\cal E}$
whose coordinates can be identified with the extensive thermodynamic variables $E^a$, with $a=1,...,n$. Then, each point of ${\cal E}$
represents a particular equilibrium state in which the system can exist. Clearly, not every point of ${\cal E}$ can be occupied by a given 
system. The set of points of ${\cal E}$ that are available to a particular system is determined by the fundamental equation $\Phi=\Phi(E^a)$,
where $\Phi$ is the thermodynamic potential \cite{callen}. Usually, $\Phi$ is identified either with the entropy $S$ or with the internal energy $U$
of the system. However, it is possible to use any thermodynamic potential, leaving the thermodynamic properties of the system unchanged.

It is possible to equip the equilibrium space with a differential geometric structure in several ways. The simplest way perhaps consists in 
introducing into ${\cal E}$ a Hessian metric 
\be
g^{H} = \frac{\partial^2 \Phi}{\partial E^a \partial E^b } d E^a d E ^b \ ,
\label{hessian}
\ee  
so that ${\cal E}$ becomes a Riemannian manifold. The line element $g^{H}$ behaves as a scalar under arbitrary changes of coordinates of ${\cal E}$, i.e., under 
diffeomorphisms $E^a \rightarrow \tilde E  ^a = \tilde E ^a (E^a)$ involving only the set of coordinates $\{E^a\}$. 

Since the function $\Phi$ can be considered as the generating function of the Hessian metric, the above geometric construction is well-defined for any particular choice of $\Phi$, 
but it does not allow to consider changes of thermodynamic potential. In fact, in classical thermodynamics a change of thermodynamic potential is defined by 
the Legendre transformation \cite{arnold} 
\be
\{\Phi, E^a, I^a\} \longrightarrow \{\tilde \Phi, \tilde E ^a, \tilde I ^ a\}\ ,
\ee
\be
 \Phi = \tilde \Phi - \delta_{kl} \tilde E ^k \tilde I ^l \ ,\quad
 E^i = - \tilde I ^ {i}, \ \  
E^j = \tilde E ^j,\quad   
 I^{i} = \tilde E ^ i , \ \
 I^j = \tilde I ^j \ ,
 \label{leg}
\ee
where $I^a$ is a set of $n$ additional variables,  $i\cup j$ is any disjoint decomposition of the set of indices $\{1,...,n\}$,
and $k,l= 1,...,i$. In particular, for $i=\{1,...,n\}$ and $i=\emptyset$, we obtain
the total Legendre transformation and the identity, respectively. The above exact definition 
shows that a Legendre transformation cannot act on an $n-$dimensional space, but on a $(2n+1)-$dimensional space ${\cal T}$ with coordinates 
$Z^A=\{\Phi, E^a, I^a\}$, with $A=0,...,2n$, which is known as the thermodynamic phase space \cite{her73}. 

The introduction of ${\cal T}$  allows us to correctly handle not only the Legendre transformations, but also the first law of thermodynamics. In fact,
in the phase space there exists a canonical contact structure determined by the fundamental Gibbs 1-form 
$\Theta = d\Phi- \delta_{ab} I^b dE^a = d\Phi - I_a dE^a$. The equilibrium space is then a subspace of ${\cal T}$ defined by the embedding map 
$\varphi: {\cal E} \longrightarrow {\cal T}$ under the condition 
\be
\varphi^* (\Theta) =0\ , \quad {\rm i.e.} \quad d\Phi = I_a d E^a \ , \quad {\rm and} \quad I_a = \frac{\partial \Phi}{\partial E^a}\ , 
\ee
where $\varphi^*$ is the pullback of $\varphi$. The above expressions can be immediately identified as the first law of thermodynamics and the equilibrium conditions, respectively. 
 Notice that the fundamental 1-form $\Theta$ is Legendre invariant in the sense that under a transformation (\ref{leg}) it transforms as $\Theta\longrightarrow \tilde \Theta =  d\tilde \Phi- \delta_{ab} \tilde I^b d\tilde E^a $. As a consequence, in the new coordinates the first law preserves its functional dependence, i. e., $d\tilde \Phi= \tilde I_a d \tilde E^a$. 

One of the goals of GTD consists in introducing a metric $g$ in ${\cal E}$ whose properties do not depend on the choice of thermodynamic potential. From the above considerations, it follows that there exists a canonical manner to introduce such a metric, namely, with $g=\varphi^*(G)$ where $G$ is a metric in ${\cal T}$ that preserves Legendre invariance. Clearly, there exist metrics $G$ that are not Legendre invariant. In particular, the flat metric $G=\delta_{AB} dZ^A d Z^B$ changes its functional dependence under a Legendre transformation, indicating that any Legendre invariant phase manifold must be curved. Furthermore, the non-degenerate metric   
\be        
G^H=(d\Phi-\delta_{ab}I^b dE^a)^2 + \delta_{ab} d E^a d I^b \ ,
\ee
whose pullback, $\varphi^*(G^H)=g^H$, generates the Hessian metric (\ref{hessian}), is not Legendre invariant. This proves that the Hessian metrics are not independent of the choice of thermodynamic potential. We have found that the most general  metric which is invariant under total Legendre transformations 
can be written as 
\begin{equation}
G = \left(d\Phi - I_a dE^a\right)^2  +\Lambda
\left(\xi_{ab}E^{a}I^{b}\right)\left(\chi_{cd}dE^{c}dI^{d}\right) \  ,
\label{gup1}
\end{equation}
where $\xi_{ab}$ and $\chi_{ab}$ are diagonal constant
tensors, and $\Lambda$  is an arbitrary Legendre invariant function of the coordinates $Z^A$. In particular, $\Lambda$ can be
chosen as an arbitrary real constant. The diagonal tensors can be expressed in terms of the 
usual Euclidean metric, $\delta_{ab}={\rm diag}(1,...,1)$, and the pseudo-Euclidean metric,  
$\eta_{ab} = {\rm diag}(-1, 1, ..., 1)$. Using additional physical conditions \cite{qstv10a}, it turns out 
that the choice 
$ \xi_{ab}=\delta_{ab}$ and $\chi_{ab}=\delta_{ab}$ 
$(\xi_{ab}=\delta_{ab}$ and $\chi_{ab}=\eta_{ab})$ 
leads to a metric which 
describes systems characterized by first (second) order phase transitions.
Moreover, the choice  
$\xi_{ab}= \left(\delta_{ab}-\eta_{ab}\right)/2$ 
allows us also to correctly handle the zero-temperature limit. We see that
Legendre invariance leaves free only the signature of $\chi_{ab}$, which can be 
fixed by the order of the phase transition under consideration.  
Since in this work we will analyze black hole configurations with 
second order phase transitions, we choose the metric  as
\begin{equation}
G^{II} =(d\Phi-\delta_{ab}I^{a}dE^{b})^{2}+ (\delta_{ab} E^{a}I^{b} )
(\eta_{cd}dE^{c}dI^{d})\ ,
\end{equation}
whose pullback generates the metric 
\begin{equation}\label{gII}
g^{II}= \left(E^a \frac{\partial \Phi}{\partial E^a}\right) \left( \eta_b^{ c} \, \frac{\partial^2\Phi}{\partial E^c \partial E^d} \, d E^b d E^d \right) 
\end{equation}
for the equilibrium manifold, where $\eta_b^{ c}={\rm diag} (-1,1,...,1)$. 
 We see that this metric can be calculated explicitly once the fundamental equation $\Phi=\Phi(E^a)$ is given.

The main point now is that the geometric properties of the equilibrium manifold ${\cal E}$ described by the metric $g^{II}$ should be related to the thermodynamic properties of the system described by the fundamental equation $\Phi(E^a)$. In particular, one expects that ${\cal E}$ be curved for systems with thermodynamic interaction and that curvature singularities in ${\cal E}$ correspond to phase transitions
of the corresponding thermodynamic system.

\section{Reissner-Nordstr\"om black hole in any dimension}
\label{sec:rn}

The solution for the charged black hole with no angular momentum (Reissner-Nordstr\"om black hole) 
can be extended to any dimension. The corresponding line element in $d$ spacetime dimensions reads \cite{ama}
\begin{equation}\label{RNmetric}
ds^2=-V\,dt^2+V^{-1}\,dr^2+r^2\,d\Omega_{(d-2)}^2\,,
\end{equation}
where $d\Omega_{(d-2)}^2$ is the line element on the $(d-2)$-dimensional unit sphere, $\Omega_{(d-2)}=2\pi^{\frac{d-1}{2} }/\Gamma(\frac{d-1}{2})$, and $V$ is defined as
\begin{equation}\label{VRN}
V=1-\frac{16\pi G M }{(d-2)\Omega_{(d-2)} }\,\frac{1}{r^{d-3}}+\frac{8\pi G }{(d-2)(d-3)}\frac{Q^2}{r^{2(d-3)}}\,.
\end{equation}

Solving the equation $V=0$, one can find the event horizon in any dimensions
and thus derive the area and the corresponding entropy.

\subsection{Thermodynamics}

The fundamental equation for the entropy reads \cite{ama}
\begin{equation}\label{entropyRNd}
S(M,Q)=\left(M+M\sqrt{1-\frac{d-2}{2(d-3)}\frac{Q^2}{M^2} }\right)^{\frac{d-2}{d-3} }\,.
\end{equation} 
Inverting (\ref{entropyRNd}), one obtains the mass function \cite{ama}
\begin{equation}\label{massRNd}
M(S,Q)=\frac{ S^{\frac{d-3}{d-2} }}{2}+\frac{d-2}{4(d-3)}\frac{Q^2}{S^{\frac{d-3}{d-2} }}\,,
\end{equation} 
that satisfies the first law of thermodynamics $dM = TdS + \phi d Q$, where $\phi$  is usually interpreted
as an electric potential. Then, the temperature and the electric potential are 
\begin{equation}\label{tempRNd}
T(S,Q)=\frac{1}{4}\,\frac{2(d-3)\, S^\frac{2(d-3)}{d-2}-(d-2)\,Q^2 }{(d-2)\,S^\frac{2d-5}{d-2} } \ , \quad
\phi(S,Q)= \frac{d-2}{2(d-3)} \frac{Q}{S^{\frac{d-2}{d-3}}}\ .
\end{equation} 
In the extremal limit,   
\begin{equation}\label{RNextremalS}
\left.\frac{Q^2}{M^2}\right |_{\textrm{extremal} }=\frac{2(d-3)}{d-2}\ 
\quad {\rm i.e.} \quad \left.\frac{Q^2}{S^\frac{2(d-3)}{d-2} }\right |_{\textrm{extremal} }=\frac{2(d-3)}{d-2}\,,
\end{equation}
the temperature of the  black hole vanishes and the electric potential is constant. Incidentally, in the extremal case, 
one gets $M^2=\phi^2 Q^2$. 
Note that this limit exists in any dimension.
We will see in the next section that the situation is different for the Kerr black hole, 
for which there is an extremal limit only up to dimension $5$ \cite{ama}.

According to Davies \cite{Davies}, phase transitions occur at those points where 
the heat capacity diverges. In general, however, according to Ehrenfest's scheme \cite{callen},
one cannot exclude the possibility that other response functions might indicate the presence of phase transitions.
We will follow this viewpoint in this work. In the case of the RN black hole,  the thermodynamic potential $M$ depends
on $Q$ and $S$ and so one can define two different response functions.
The heat capacity at constant $Q$ reads
\begin{equation}\label{CQRNd}
C_Q=\frac{M_S}{M_{SS}}=-\frac{(d-2)S\left((d-2)\,Q^2-2(d-3)\, S^\frac{2(d-3)}{d-2}\right) }{(d-2)(2d-5)\,Q^2-2(d-3)\, S^\frac{2(d-3)}{d-2} }\ ,
\end{equation} 
where $M_S = \partial M/\partial S$, etc. Moreover, 
in this ensemble one can also consider the  isentropic compressibility
\begin{equation}\label{KSRNd}
\kappa_S=\frac{1}{Q M_{QQ}}=\frac{2(d-3)}{d-2}\frac{S^\frac{d-3}{d-2} }{Q}\ .
\end{equation}

We note that the only possible divergence is that of the heat capacity, which takes place when the denominator of
(\ref{CQRNd}) is zero, i.e., whenever 
\begin{equation}\label{daviespoint}
\left.\frac{Q^2}{S^{\frac{2(d-3)}{d-2} }}\right |_{\textrm{phase transition} }=\frac{2(d-3)}{(2d-5)(d-2)}\,.
\end{equation} 
One can prove that this value is in the black hole region, i.e., that the condition 
\begin{equation}
\frac{Q^2}{M^2} \leq \frac{2(d-3)}{d-2}
\label{rnbhr}
\end{equation}
is  satisfied. 
{In fact, by} using Eq. (\ref{entropyRNd}), we can rewrite Eq. (\ref{daviespoint}) as 
{\begin{equation}\label{daviespoint2}
{\left.\frac{Q^2}{ M^2}\right|_{\textrm{phase transition}}=\frac{2(d-3)(2d-5)}{(d-2)^3}\,,}
\end{equation}
which is easily proven to be inside the black hole region for any value of $d>3$.

It is interesting to note that the phase transition structure of black holes can depend on the chosen ensemble. 
For instance, if we use the ensemble corresponding to the ``enthalpy", $H=M-\phi Q$, 
\begin{equation}\label{HRN}
H(S,\phi)=-S^{\frac{d-3}{d-2}}\frac{2\phi^2\,(d-3)-d+2}{2\,(d-2)}\,,
\end{equation} 
from which we can calculate 
\begin{equation}\label{CphiRN}
C_\phi=\frac{H_S}{H_{SS}}=-(d-2)\,S\ ,
\end{equation}
we observe that the heat capacity at constant $\phi$ has no singularities; hence we expect no phase transitions from the thermodynamic analysis in this ensemble.

\subsection{Geometrothermodynamics}
Given the fundamental equations (\ref{entropyRNd}) and (\ref{massRNd}), and the general metric (\ref{gII}),
we can calculate the particular metric and the scalar curvature for the RN black hole, both in the entropy and
in the energy representations.
The metric with $\Phi=S$ and $E^a=\{M,Q\}$ reads
\begin{equation}\label{gIIRNdS}
\begin{split}
g^{II}_S\,=\,& \frac{\left[\frac{1}{2}M(2+\mathcal{E}(M,Q))\right]^\frac{2(d-2)}{d-3}}{\mathcal{D}(M,Q) }
\bigg\{  -4(d-2)^2(d-3)\bigg[\big[(d-2)^2Q^2-4(d-3)M^2\big]\mathcal{E}(M,Q)\\
&\left.\left.+\,2(d-1)(d-2)Q^2-8(d-3)M^2\bigg]\, dM^2\right.\right.\\
&-\,2(d-2)^3\bigg[\big[(d-2)Q^2-2(d-3)^2M^2\big]\mathcal{E}(M,Q)-4(d-3)^2M^2\bigg]\, {dQ^2}\bigg\}\ ,
\end{split}
\end{equation}
where
\begin{equation}
\mathcal{D}(M,Q)=M^2(d-3)^4\bigg((d-2)Q^2-2(d-3)M^2\bigg) \mathcal{E}(M,Q) \bigg(2+\mathcal{E}(M,Q)\bigg)^2
\end{equation}
and
\begin{equation}
\mathcal{E}(M,Q)=\sqrt{4-2\frac{(d-2)}{d-3}\frac{Q^2}{M^2} }\,.
\end{equation}
The scalar curvature is 
\begin{equation}\label{curvRNdS}
R^{II}_S\,=\frac{\mathcal{N}_1(M,Q)}{\mathcal{A}_1(M,Q)^2\mathcal{B}_1(M,Q)^2 }\,,
\end{equation}
where
\begin{equation}
{\mathcal{A}_1(M,Q)=\,\big[(d-2)^2Q^2-4(d-3)M^2\big]\mathcal{E}(M,Q)+\,2(d-1)(d-2)Q^2-8(d-3)M^2}
\end{equation}
and
\begin{equation}
{\mathcal{B}_1(M,Q)=\big[(d-2)Q^2-2(d-3)^2M^2\big]\mathcal{E}(M,Q)-4(d-3)^2M^2\,,}
\end{equation}
{which are proportional to the metric components in (\ref{gIIRNdS}).}

Using a software for algebraic manipulations, we find that the only real {positive} root of the denominator of the curvature scalar (\ref{curvRNdS}) is given {by solving
$\mathcal{A}_1(M,Q)=0$ and gives 
\begin{equation}\label{singpoint}
\left.M\right|_{\textrm{singularity}}=\frac{\sqrt 2}{2}\frac{(d-2)^2\,Q}{\sqrt{d-3}\,\sqrt{d-2}\,\sqrt{2d-5}}\,,
\end{equation}
 which  is easily seen to be equivalent to (\ref{daviespoint2}).}
Furthermore, it can be proven that $\mathcal{N}_1(M,Q)$ is never zero where $\mathcal{A}_1(M,Q)$ and $\mathcal{B}_1(M,Q)$ are. 
From Eqs. {(\ref{daviespoint2})} and (\ref{singpoint}), we can conclude that in fact the curvature singularities are located exactly at those points where phase transitions occur.

To show that the above results are invariant, we now use as thermodynamic potential $\Phi=M$ and $E^a=\{S,Q\}$, satisfying the fundamental equation (\ref{massRNd}).  Then, from the general 
thermodynamic metric (\ref{gII}), we obtain the metric 
{\begin{equation}\label{gIIRNdU}
\begin{split}
g^{II}_M\,=\,&\frac{\mathcal{A}_2(S,Q)\,S^{-\frac{2(d-3)}{d-2}}}{d-3}\bigg\{-\frac{\mathcal{B}_2(S,Q)}{16(d-2)^3S^2}\,dS^2
+\frac{1}{8(d-3)}\,dQ^2\bigg\}\ ,
\end{split}
\end{equation}}
from which we compute the curvature scalar 
\begin{equation}\label{curvRNdU}
{R^{II}_M\,=\frac{\mathcal{N}_2(S,Q)}{\mathcal{A}_2(S,Q)^7\mathcal{B}_2(S,Q)^3 }\,,}
\end{equation}
where
\begin{equation}
\mathcal{A}_2(S,Q)=2(d-3)^2S^\frac{2(d-3)}{d-2}+(d-1)(d-2)Q^2 
\end{equation}
and
{\begin{equation}
\mathcal{B}_2(S,Q)=\left( 2d-5 \right)  \left(d-2 \right) {Q}^{2}-2\left(d-3
 \right) {S}^{{\frac {2(d-3)}{d-2}}}	\,.
\end{equation}}

Using again a software for algebraic manipulations, one can see that the only points of divergence of $R^{II}_M$ are given by
{\begin{equation}\label{singpoint}
\left.\frac{Q^2}{S^{\frac{2(d-3)}{d-2} }}\right |_{\textrm{singularity} }=\frac{2(d-3)}{(2d-5)(d-2)}\,,
\end{equation}}
which coincides with the condition for the phase transitions (\ref{daviespoint}).
Once more we see a concrete relationship between the curvature of the metric (\ref{gII}) and 
the thermodynamic interaction.

To consider the behavior of the GTD analysis with respect to different ensembles, we turn now back to  the general metric (\ref{gII}) and write it with the fundamental equation (\ref{HRN}) so that $\Phi=H$ and $E^a=\{S,\phi\}$. The resulting metric reads
\begin{equation}\label{gIIRNdH}
\begin{split}
g^{II}_H\,=\,&\frac{(d-3)^2(6\phi^2d-14\phi^2-d+2)}{4(d-2)^3}\bigg\{\frac{
2\phi^2d-6\phi^2-d+2}{S^{\frac{2}{d-2}}(d-2)^2}\,dS^2
+4S^{2\frac{d-3}{d-2}}\,d\phi^2\bigg\}\,.
\end{split}
\end{equation}
Consequently, the scalar curvature reads
\begin{equation}\label{curvRNdH}
R^{II}_H\,=\frac{\mathcal{N}_3(S,\phi)}{\mathcal{A}_3(S,\phi)^3\mathcal{B}_3(S,\phi)^2 }\,,
\end{equation}
where
\begin{equation}
\mathcal{A}_3(S,\phi)=6\phi^2d-14\phi^2-d+2\,,
\end{equation}
i.e. the conformal factor in the metric (\ref{gIIRNdH})
and 
\begin{equation}
\mathcal{B}_3(S,\phi)=2\phi^2\,(d-3)-d+2\,.
\end{equation}
Hence the first factor in the denominator, being the conformal factor in the metric (\ref{gIIRNdH}), is equal to $SH_S+\phi H_{\phi}$, which, in turn, according to Euler's identity, is proportional to $H$.  Thus, the first term in the denominator of the scalar curvature is zero if and only if the thermodynamic potential vanishes, $H=0$. 
Considering the equation of state $\phi=\partial M/\partial Q$, the second factor turns out to be zero for $S^2=[Q^2(d-2)/2(d-3)]^{(d-2)/(d-3)}$, which corresponds exactly to the extremal black hole limit (\ref{RNextremalS}) with zero temperature. This is due to the fact that in this case the metric $g^{II}_H$ becomes degenerate in the extremal limit. Thus, the only singularities arise from the limits of applicability of the thermodynamic approach to black holes, where we also expect the GTD approach to break down.

We  conclude that the scalar curvature in this ensemble has no true singularities, signaling the absence of phase transitions, in agreement with the results obtained from the study of the corresponding heat capacity (\ref{CphiRN}).

As a final remark, we point out that the Hessian metric (\ref{hessian}) with $\Phi=M$ (Weinhold's metric) for this case is curved, but the analysis of its scalar curvature gives no special information
about the phase transition points. Moreover, for $\Phi=S$ (Ruppeiner's metric) the equilibrium space is flat \cite{ama}. In 
\cite{gpw11}, this last result is interpreted as an indication that a divergence of the heat capacity does not represent a 
phase transition.

\section{Kerr black hole in any dimension}
\label{sec:kerr}

The solution for the Kerr black hole in arbitrary dimension corresponds to taking a Myers-Perry
black hole with only one angular momentum different from zero. The line element is \cite{EmparanMyers}
\begin{equation}\label{Kerrmetric}
\begin{split}
ds^2=&-dt^2+\frac{\mu }{r^{d-5}\rho^2}\,(dt+a\,{\rm sin}^2\theta\,d\varphi)^2+\frac{\rho^2}{\Delta}\,dr^2+\rho^2\,d\theta^2+(r^2+a^2){\rm sin}^2\theta\,d\varphi^2\\
&+r^2{\rm cos}^2\theta\,d\Omega_{(d-4)}^2\,,
\end{split}
\end{equation}
where
\begin{equation}\label{kerrquantities}
\rho^2=r^2+a^2{\rm cos}^2\theta\,, \qquad \Delta=r^2+a^2-\frac{\mu }{r^{d-5}}
\end{equation}
and the physical mass $M$ and angular momentum $J$ are related to $\mu$ and $a$ by
\begin{equation}\label{physicalKerrparam}
M=\frac{(d-2)\partial\Omega_{d-2} }{16\pi G }\,\mu\,, \qquad J=\frac{2}{d-2}\,M\,a\,,
\end{equation}
where $\partial\Omega_{d-2}$ is the area of the $(d-2)$-dimensional unit sphere.

\subsection{Thermodynamics}

\noindent Following the notation of \cite{ama}, the fundamental equation for the mass is
\begin{equation}\label{massKerrd}
M(S,J)= \frac{d-2}{4} S^{\frac{d-3}{d-2} }\left(1+\frac{4J^2}{S^2}\right)^\frac{1}{d-2}
\end{equation} 
In the general case, it is not possible to invert this equation, so we will work with the mass representation only. 
From (\ref{massKerrd}) it follows that the temperature $T=\frac{\partial M }{\partial S }$ and the angular velocity at the horizon
$\Omega=\frac{\partial M }{\partial J }$ are 
\begin{equation}\label{tempKerrd}
T(S,J)=\frac{(d-3)\left(1+4\frac{d-5}{d-3}\frac{J^2}{S^2}\right) }{4S^\frac{1}{d-2}\left(1+4\frac{J^2}{S^2}\right)^\frac{d-3}{d-2} }\ ,
\quad 
\Omega(S,J) =  \frac{2 J}{S^\frac{d-1}{d-2}\left(1+4\frac{J^2}{S^2}\right)^\frac{d-3}{d-2}}\ ,
\end{equation} 
from which it can be easily seen that an extremal limit for the Kerr black hole exists only for $d\leq5$.
When $d=4$ the limit is the usual Kerr bound $J/M^2=1$ and for $d=5$ it is the Myers-Perry 
black hole bound $J^2/M^3=16/27$ (see e.g. \cite{ama}).

To investigate the phase transition structure of this black hole, we calculate the 
corresponding response functions in this representation. The heat capacity at constant angular momentum $J$ reads 
\begin{equation}\label{CJKerrd}
C_J=\frac{M_S}{M_{SS}}=-\frac{(d-2)S(S^2+4J^2)\big[(d-3)S^2+4(d-5)J^2\big] }{48(d-5)J^4-24S^2J^2+(d-3)S^4} \ ,
\end{equation} 
and the isentropic compressibility can be expressed as
\begin{equation}\label{KSKerrd}
\kappa_S=\frac{1}{J\,M_{JJ}} =-\frac{1}{2\,J\, S^\frac{d-5}{d-2} }\cdot\frac{(d-2)\left(S^2+4J^2\right)^\frac{2d-5}{d-2} }{4(d-4)J^2-(d-2)S^2}\ .
\end{equation} 

We note that in this case there can be divergences in both response functions. 
First of all let us focus on the heat capacity.
The divergences in this case are situated at the points which satisfy the relation
\begin{equation}\label{daviespointK}
\left.\frac{J^2}{S^2}\right|_{\textrm{phase transition} }=\frac{1}{4}\frac{d-3}{3+\sqrt{-3d^2+24d-36} }\,.
\end{equation} 
The r.h.s. of Eq.(\ref{daviespointK}) is real only if $d=4,5,6$, so that we can have a divergence in the heat capacity 
(a phase transition \`a la Davies) only for $d=4,5,6$.
Furthermore, the isentropic compressibility diverges when
\begin{equation}\label{singKK}
\left.\frac{J^2}{S^2}\right|_{\textrm{phase transition} }=\frac{1}{4}\frac{d-2}{d-4} \ ,
\end{equation}
under the condition that $d>4$.

For $d=4,5$, we have to check that the singularities of the heat capacity are in the black hole region.
Indeed, this can easily be checked. 
For example, for $d=4$, we can obtain $S$ from Eq.(\ref{daviespointK}) and evalute $J/M^2$ at this critical value. Since 
the result is $\left.J/M^2\right|_{Scritical}=\sqrt{3+2\sqrt3}/(2+\sqrt3)$, which is less than the extremal limit $J/M^2=1$, 
we can affirm that  this point belongs to the black hole region. A similar analysis can be performed for $d=5$ to
prove that all the singularities of the response functions are in the black hole region.

\subsection{Geometrothermodynamics}

Given the fundamental equation (\ref{massKerrd}) and the general metric (\ref{gII}),
we can calculate the particular metric and the scalar curvature for the Kerr black hole only in the mass representation.
Fortunately, in \cite{Termometrica}, we have proven a conformal relation which enables us to write $g^{II}_S$
in terms of $g^{II}_M$, the relation being for the case of the Kerr black holes
\begin{equation}\label{conformalrel}
g^{II}_S=-\frac{M-J \, \Omega}{T^2(S\,T+J\,\Omega)}\,g^{II}_M\,,
\end{equation}
where $T$ is the temperature and $\Omega$ is the angular velocity at the horizon, as usual.
First we calculate (\ref{gII}) in the mass representation. Using eq. (\ref{massKerrd}) the metric reads
\begin{equation}\label{gIIKerrdU}
\begin{split}
g^{II}_M\,=\,&\frac{(d-3)}{16(d-2)\left(S^2+4J^2\right)^\frac{2(d-3)}{d-2} }\bigg\{S^\frac{-6}{d-2}\bigg[48(d-5)J^4-24S^2J^2+(d-3)S^4\bigg]dS^2\\
&-8S^\frac{2(d-5)}{d-2}\bigg[4(d-4)J^2-(d-2)S^2\bigg] dJ^2\bigg\}\,.
\end{split}
\end{equation}
With the use of eqs. (\ref{conformalrel}) and (\ref{gIIKerrdU}), the metric in the entropy representation then reads
{\begin{equation}\label{gIIKerrdS}
\begin{split}
g^{II}_S\,=\,&-\frac{16\,S^\frac{6}{d-2}\left(S^2+4J^2\right)^\frac{d-4}{d-2}\big[4(d-4)J^2+(d-2)S^2\big] }{(d-3)\bigg[(d-3)S^2+4(d-5)J^2\bigg]^2 }\cdot g^{II}_M\,.
\end{split}
\end{equation}
}
The scalar curvature in both cases can be calculated. It turns out to be 
\begin{equation}\label{curvKerrdU}
R^{II}_M\,=\frac{\mathcal{N}_4(S,J)}{(d-3)\mathcal{A}_4(S,J)^2\mathcal{B}_4(S,J)^2 }\,,
\end{equation}
where
\begin{equation}\label{coeffA3}
\begin{split}
\mathcal{A}_4(S,J)=48(d-5)J^4-24J^2S^2+(d-3)S^4
\end{split}
\end{equation}
and
\begin{equation}\label{coeffB3}
\mathcal{B}_4(S,J)=4(d-4)J^2-(d-2)S^2\,,
\end{equation}
while 
\begin{equation}\label{curvKerrdS}
R^{II}_S\,=\frac{\mathcal{N}_5(S,J)}{\mathcal{A}_4(S,J)^2\mathcal{B}_4(S,J)^2\mathcal{C}_5(S,J)^3 }\,,
\end{equation}
where $\mathcal{A}_4(S,J)$ and $\mathcal{B}_4(S,J)$ are the same as in (\ref{coeffA3}) and (\ref{coeffB3}), and
\begin{equation}
\mathcal{C}_5(S,J)=4(d-4)J^2+(d-2)S^2
\end{equation}
{is proportional to the conformal factor in (\ref{gIIKerrdS}).}
First we note that $\mathcal{C}_5(S,J)$ is always different from zero. 
Afterwards, it is immediate to see that $\mathcal{A}_4(S,J)$ is exactly the denominator of the heat capacity  (\ref{CJKerrd})
and $\mathcal{B}_4(S,J)$ is the denominator of the compressibility  (\ref{KSKerrd}). Since both of them are in the 
denominators of $R^{II}_M$ and $R^{II}_S$ and since the numerators do not vanish at the points of singularity, 
we conclude that again we have a concrete relationship between the singularities of the curvature of the metric 
(\ref{gII}) and the second order phase transition structure.


\section{Conclusions}
\label{sec:con}

In this work, we analyzed the geometric structure of the equilibrium manifold of higher dimensional black holes.
To this end, we used the approach of GTD, a formalism that represents the thermodynamic properties, like interaction and
phase transitions, in terms of concepts of differential geometry, like curvature and singularities, in a way that resembles the geometric interpretation of field theories. 

We analyzed two of the most interesting higher dimensional black hole configurations, namely, the Reissner-Nordstr\"om and the Kerr black holes. First, we derive all the critical points that follow from the analysis of the divergencies of the thermodynamic response functions. In black hole thermodynamics, the critical points of the heat capacity are usually associated with the occurrence of second order phase transitions.  Here we analyzed the divergencies of all the response functions 
and showed that GTD reproduces the behavior near the critical points.

In the case of the Reissner-Nordstr\"om black hole, we found that, if we use the ensemble associated with the mass of the black hole, there exists only one phase transition of second order. On the other hand, if we use the ensemble associated with the enthalpy, no phase transitions exist. This is in accordance with the well-known result that the phase transition structure of black holes can depend on the ensemble. We then explored the geometric properties of the corresponding equilibrium space by using GTD, with the mass as thermodynamic potential, and found that a curvature singularity appears exactly at that point where the phase transition occur. If, instead, we use the enthalpy as thermodynamic potential, GTD provides a singularity-free equilibrium manifold. Thus, GTD reproduces correctly the thermodynamic phase transition structure of the Reissner-Norstr\"om black hole.

In the case of the higher dimensional Kerr black hole, the response functions predict more phase transitions than in the Reissner-Nordstr\"om case. It is not possible to compute explicitly other thermodynamic potentials, and so we perform all the calculations in the mass and entropy representations. Again, we found true curvature singularities at the same points where phase transitions take place. This result reinforces the conclusion that GTD is able to correctly reproduce the phase transition structure of black holes.


\begin{thebibliography}{99}

\bibitem{frankel} T. Frankel, {\em The geometry of physics: An introduction} (Cambridge University Press, Cambridge, UK, 1997).


\bibitem{gibbs} J. W. Gibbs, {\it The collected works}, Vol. 1, Thermodynamics (Yale 
University Press, 1948).



\bibitem{car} C. Caratheodory, {\it Untersuchungen \"uber die Grundlagen der 
Thermodynamik},   Math. Ann. {\bf 67}, 355 (1909).  

\bibitem{rao45} C. R. Rao, Bull. Calcutta Math. Soc. {\bf 37}, 81 (1945).

\bibitem{amari85} S. Amari, {\it Differential-Geometrical Methods in Statistics} (Springer-Verlag, Berlin, 1985).


\bibitem{am} J. E. \AA man, I. Bengtsson, and N. Pidokrajt, Gen. Rel. Grav.
\textbf{35} 1733 (2003).

\bibitem{ama} J. E. \AA man and N. Pidokrajt, Phys. Rev. D \textbf{73}, 024017
(2006).

\bibitem{aman06a} J. E.  \AA man and N. Pidokrajt, Gen. Rel. Grav.\textbf{38},
1305 (2006).


\bibitem{scws} J. Shen, R. G. Cai, B. Wang, and R. K. Su, [gr-qc/0512035].

\bibitem{caicho99} R. G. Cai and J. H. Cho, Phys. Rev. D \textbf{60}, 067502 (1999).



\bibitem{sst} T. Sarkar, G. Sengupta, and B. N. Tiwari, J. High Energy Phys.
\textbf{0611} 015 (2006).

\bibitem{med} A. J. M. Medved, Mod. Phys. Lett. A \textbf{23}, 2149 (2008).

\bibitem{mz} B. Mirza and M. Zamaninasab,  JHEP, 0706:059 (2007).    

\bibitem{hernando2} H. Quevedo, Gen. Rel. Grav. {\bf 40}, 971 (2008).


\bibitem{callen}
H. B. Callen, {\it   "Thermodynamics and an Introduction to Thermostatics".}
(John Wiley and Sons, Inc., New York, 1985).


\bibitem{quev07} H. Quevedo, J. Math. Phys.
{\bf 48}, 013506 (2007).

\bibitem{her73} R. Hermann, {\it Geometry, physics and systems} (Marcel
Dekker, New York, 1973).



\bibitem{qstv10a} H. Quevedo, A. S\'anchez, S. Taj, and A. V\'azquez, Gen. Rel. Grav. {\bf 43} (2011) 1153. 


\bibitem{liv-rev}
R. Emparan and H. S. Reall,
Living Rev. Rel. {\bf 11}, 6 (2008).


\bibitem{gpw11} 
L. Gergely, N. Pidokrajt and S. Winitzki,
 Eur. Phys. J. C 71:1569 (2011).


\bibitem{roy11a}
R. Banerjee and D. Roychowdhury, JHEP 1111:004 (2011). 

\bibitem{roy12}
R. Banerjee, S. K. Modak and D. Roychowdhury, JHEP 1210:125 (2012). 

\bibitem{roy11b}
R. Banerjee, S.  Ghosh and D. Roychowdhury, Phys. Lett. B {\bf 696} 156 (2011). 

\bibitem{qs08} H. Quevedo and A. S\'anchez, JHEP 09:034 (2008).

\bibitem{konzhi09}
  R.~A.~Konoplya and A.~Zhidenko,
  Phys.\ Rev.\ Lett.\  {\bf 103}, 161101 (2009)
  



\bibitem{arnold} V. I. Arnold, {\it Mathematical Methods of Classical Mechanics}
(Springer Verlag, New York, 1980).







\bibitem{Davies}
P. C. W. Davies, 
Rep. Prog. Phys. \textbf{41} 1313 (1978).



\bibitem{EmparanMyers} R. Emparan, R. C. Myers, JHEP {\bf 09}, 025 (2003).


\bibitem{Termometrica}
A. Bravetti, F. Nettel, {\it Second order phase transitions and thermodynamic geometry: a general approach}, 
arXiv:1208.0399


\end{thebibliography}
\end{document}